\documentclass[twocolumn,journal]{IEEEtran}

\ifCLASSINFOpdf
  \usepackage[pdftex]{graphicx}
  \DeclareGraphicsExtensions{.pdf,.jpeg,.png}
\else
  \usepackage[dvips]{graphicx}
\fi

\usepackage{multirow}
\usepackage{makecell}
\usepackage{amsmath,amssymb}
\allowdisplaybreaks[4]
\usepackage{epstopdf}
\usepackage{epsfig}
\usepackage{cite}
\usepackage{subcaption}
\usepackage{CJKutf8}
\usepackage{color}
\usepackage{multicol}
\usepackage{stfloats} 
\usepackage{graphicx}
\usepackage{float}
\usepackage{bm}

\usepackage{booktabs}

\usepackage{hyperref} 
\hypersetup{
	colorlinks=true,       
	linkcolor=red,         
	anchorcolor=black,     
	citecolor=blue,        
	urlcolor=blue,         
}
\usepackage{mfirstuc}
\usepackage[switch]{lineno} 
\usepackage[linesnumbered, ruled]{algorithm2e}

\usepackage{algpseudocode} 

\newtheorem{theorem}{Theorem}

\UseRawInputEncoding
\begin{document}
\title{
	JSSAnet: Theory-Guided Subchannel Partitioning and Joint Spatial Attention for Near-Field Channel Estimation
}

\author{Zhiming~Zhu,~
	Shu~Xu,~\IEEEmembership{Student Member,~IEEE},
	Chunguo~Li,~\IEEEmembership{Senior Member,~IEEE},
	Yongming~Huang,~\IEEEmembership{Fellow,~IEEE},
	and
	Luxi~Yang,~\IEEEmembership{Senior Member,~IEEE}
}

\maketitle

\renewcommand{\thefootnote}{}
\footnotetext{
	
	Z.~Zhu, S.~Xu, Y.~Huang and L.~Yang are with the National Mobile Communications Research Laboratory, School of Information Science and Engineering, Southeast University, Nanjing 210096, China, and also with Purple Mountain Laboratories, Nanjing 211111, China (e-mail:\{zhuzm,~shuxu,~huangym,~lxyang\}@seu.edu.cn).
	
	C.~Li is with the National Mobile Communications Research Laboratory,
	School of Information Science and Engineering, Southeast University, Nanjing 210096, China (e-mail: chunguoli@seu.edu.cn).
	}

\begin{abstract}
	The deployment of extremely large-scale antenna array (ELAA) in sixth-generation (6G) communication systems introduces unique challenges for efficient near-field channel estimation.
	To tackle these issues, this paper presents a theory-guided approach that incorporates angular information into an attention-based estimation framework. 
	A piecewise Fourier representation is proposed to implicitly encode the near-field channel’s inherent nonlinearity, enabling the entire channel to be segmented into multiple subchannels, each mapped to the angular domain via the discrete Fourier transform (DFT).
	Then, we develop a joint subchannel-spatial-attention network (JSSAnet) to extract the spatial features of both intra- and inter-subchannels. 
	To guide theoretically the design of the joint attention mechanism, we derive upper and lower bounds based on approximation criteria and DFT quantization loss mitigation, respectively.
	Following by both bounds, a JSSA layer of an attention block is constructed to assign independent and adaptive spatial attention weights to each subchannel in parallel.
	Subsequently, a feed-forward network (FFN) of an attention block further captures and refines the residual nonlinear dependencies across subchannels.
	Moreover, the proposed JSSA map is linearly computed via element-wise product combining large-kernel convolutions (DLKC), maintaining strong contextual learning capability.
	Numerical results verify the effectiveness of embedding sparsity information into the attention network and demonstrate JSSAnet achieves superior estimation performance compared with existing methods.
	
\end{abstract}

\renewcommand{\thefootnote}{\arabic{footnote}}

\section{introduction}
\IEEEPARstart{T}{he} sixth-generation (6G) communication systems are expected to meet growing demands, such as the higher spectral efficiency (SE) and enhanced spatial degrees of freedom (DoF)\cite{You2025_Next,Wang2023_6Gservery}.
Meanwhile, extremely large-scale multiple-input-multiple-output (XL-MIMO) is regarded as a promising paradigm to advance conventional MIMO technologies, where extremely large-scale array (ELAA) is deployed to meet the expected demands of 6G \cite{Cui2023_survey,Wang2024_Tutorial}.

Compared with conventional massive MIMO, the deployment of ELAA involves not only a significantly larger array aperture but also arises from a fundamental changes in electromagnetic (EM) characteristics in the communication systems \cite{Selvan2017,Zhang2022}.
The EM propagation field is divided into near-field and far-field regions by the Rayleigh distance $d_R = \frac{2D^2}{\lambda}$ \cite{Sun2025_How}, that is determined by the array aperture $D$ and signal wavelength $\lambda$.
The extended Rayleigh distance of XL-MIMO increases the likelihood that the communication link builds in the near-field region.
Unlike the planar wavefronts in far-field communication, near-field propagation is characterized by spherical wavefronts impinging on the antenna array.
Consequently, the large number of antenna elements, combined with the paradigm shift in channel models, brings new challenges for near-field channel estimation \cite{Cui2023_survey,Wang2024_Tutorial}.
The expanded antenna aperture increases the size of the channel matrix, leading to the higher computational complexity in channel estimation.
Spherical wavefronts introduce additional channel coefficients that need to be estimated.
In recent years, near-field channel estimation has received significant attention.

Classically, least squares (LS) and minimum mean squared error (MMSE) estimators are directly applied to near-field channel estimation without requiring channel model \cite{mimo_founfations}.
However, LS comes from a linear formulation of channel reconstruction and is sensitive highly to noise.
Meanwhile, MMSE adopts channel statistical characteristics to generate a weight matrix to minimize the estimated errors of LS solutions.
However, it is difficult to prepare the knowledge of channel correlation matrix, especially the ELAA brings the higher computation of matrix multiplication.

Inspired by far-field channel estimation techniques, some approaches exploit the low-rank property of near-field channel matrices and employ the compressive sensing (CS) algorithms to estimate channel parameters.
In conventional channel estimation frameworks, the original channel matrix can be presented as a sparse angular representation by discrete Fourier transform (DFT) \cite{Lee_CEomp}.
However, due to significant angular spread and beam squint in near-field channels, DFT-based representations become ineffective for accurate channel estimation \cite{Zhang2022_Beam,Cui2023_Field,zhu_twc}.
To explore the near-field channel sparsity, studies in \cite{Han2020_Channel, NearCE_Dai} propose that the near-field channel is projected into the sparse spatial location domain by the transform dictionaries.
Subsequently, the orthogonal matching pursuit (OMP) algorithms are applied to estimate sparse channel coefficients and reconstruct the near-field channel.
In \cite{Han2020_Channel}, the near-field transform matrix is constructed in the Cartesian domain by uniformly discretizing the communication coverage area into grids within the $x\text{-}y$ plane.
Alternatively, the authors of \cite{NearCE_Dai} propose the near-field channel is projected into the polar domain by discretizing both the distance and angle grids.
Moreover, the authors of \cite{Liu2025_Sensing} introduce a novel discrete prolate spheroidal sequence (DPSS)-based eigen-dictionary for improving near-field channel estimation.
As a critical implementation in the XL-MIMO, reconfigurable intelligent surface (RIS)-assisted channels are modeled under near-field conditions.
In \cite{Wei2022_Codebook}, the authors proposed extending the near-field RIS-aided channel dictionary is extended from two-dimensional (2D) to three-dimensional (3D) space in the $x\text{-}y\text{-}z$ Cartesian coordinate system.
With polar-based dictionaries, the sparse Bayesian learning (SBL) is employed to reconstruct parameters of the RIS channel in \cite{Yu2023_Channel}.
However, these approaches incur the high computational overhead caused by the large-dimensional dictionary.
To address this issue, an efficient damped Newtonized OMP algorithm with hierarchical dictionaries in \cite{Lu2024_Field} is proposed to refine the path coefficients layer in a layer-wise manner.
Meanwhile, the works in \cite{zhu_twc} demonstrates that near-field channels exhibit angular sparsity and proposes an adaptive joint SBL algorithm that alternately refines angular and distance information for near-field orthogonal frequency division multiplexing (OFDM) channel reconstruction.
Reference \cite{Pan2023_RIS} designs a downsampled Toeplitz covariance matrix to decouple the angle and distance, followed by one-dimensional (1D) spatial search to joint estimate spatial angle-distance information of RIS-assisted channels.

CS-based channel estimation algorithms share a common limitation, namely that estimation performance depends on the dictionary choice and stopping criterion.
As an alternative choice, the machine learning (ML)-based channel estimation techniques has consequently been developed \cite{Sun2019_Application, Wang2020_Thirty}.
In channel estimation tasks, the convolutional neural network (CNN) is employed to explore the spatial and frequency correlation of channels in \cite{Dong2019_CNN_CE}.
The authors of \cite{Liu2022_Deep} frame the channel estimation problem as an image denoising task, utilizing residual CNN to refine the coarse LS-based estimates.
Motivated by channel sparsity, CNNs incorporating channel sparsity are proposed to enhance the denoising performance and robustness in \cite{Jiang2021_Dual, Xu2024_Deep}, where the coarse LS-based estimates and their sparse angular representations are simultaneously fed into a dual CNN (DuCNN) architecture.
However, deep learning (DL)-based near-field channel estimation is still in its early stages.
The works of \cite{Lee2022_Intelligent} is exploiting CNN to extract spatial correlation of near-field channels for obtaining channel coefficients.
The model-assisted DL framework, called the learned iterative shrinkage and thresholding (LISTA), is proposed for near-field channel estimation in \cite{Zhang2023_Field}, where the near-field transform dictionary is embedded into the deep neural network (DNN) in the ISTA procedure.
To capture the nonlinear relationship of channel parameters, \cite{Yuan2025_Neural} proposes the DL-assisted scheme where the recovery problem is formulated in a probabilistic form.
To circumvent the exhaustive search in \cite{Han2020_Channel}, \cite{Li2025_Keypoint} treats near-field channel sparse representations as images and develops a keypoint detection network to estimate channel spatial information.
In \cite{Wang2025_Field}, the polar-based dictionary is employed in the ISTA procedure to improve the near-field channel sparse reconstruction.
Meanwhile, the transformer excels at the in-context learning (ICL) capabilities due to self-attention mechanisms \cite{Liang2017_Why,Guo2023_How}.
Although the self-attention mechanism demonstrate superior performance in conventional channel estimation compared to other DL schemes \cite{Fan2024_Spatial,Luan2023_Channelformer}, the quadratic computational complexity of self-attention maps is impractical for the expanded near-field channel matrix.
The study of \cite{zhu_tcom} identifies CNN's poor adaptation to near-field non-stationarity, and the proposed spatial attention network (SAN) with linear complexity substantially outperforms CNN-based approaches.
However, this approach fails to exploit the intrinsic sparse properties of near-field channels.

In this work, we focus on developing a joint attention-based near-field channel estimation framework that explicitly incorporates spatial sparsity.
Specifically, the main contributions of this work are as follows:
\begin{itemize}
	\item We formulate near-field channel estimation as a denoising task on the LS-based coarse estimated channel.
	Inspired by \cite{Jiang2021_Dual, Xu2024_Deep, zhu_tcom}, we propose a hybrid strategy that embeds sparse angular information into an attention-based neural network to enhance near-field channel estimation.
	To avoid using expanded near-field dictionaries, the piecewise Fourier representation is introduced to implicitly and approximately encode the nonlinearity in the near-field channels utilizing the diverse spatial angles of subarrays.

	\item Building on the piecewise Fourier representation, the channel is partitioned into subchannels, each of which is projected into angular domain via DFT.
	For theoretically guiding the design of the joint attention mechanism,
	we derive a theoretical lower bound on the piecewise number to satisfy the approximation criterion between the piecewise Fourier vector and near-field array response vector.
	In addition, we establish a theoretical upper bound to ensure that the angular diversity among subchannels is maintained, thereby preserving critical spatial information after DFT.

	\item We propose a joint subchannel-spatial-attention network (JSSAnet) for near-field channel estimation.
	Cater to the channel partitioning, the design of attention blocks follows a decoupling-fusion strategy.
	First, a JSSA layer provides the personalized and independent spatial attentions for each subchannel in parallel.
	Second, a forward-feed network (FFN) layer is constructed to fuse extracted features across subchannels and model the nonlinear representations of near-field channel.
	Notably, our JSSA map is generated by the element-wise product combining decomposed large kernel convolutions (DLKC) for a linear computational complexity without compromising ICL capability.
	
	\item Numerical results demonstrate both the effectiveness of ELAA partitioning and the superior performance of our JSSAnet compared to other near-field channel estimation schemes.
	Ablation studies verify that jointly focusing on subchannel spatial sparsity significantly enhances the accuracy of near-field channel estimation.
\end{itemize}

The remainder of this paper is organized as follows.
Section \ref{section:system model} introduces the near-field OFDM channel model and formulates the channel denoising task.
Next, \ref{section:subchannel} present the representation of piecewise Fourier vector for the steering vector.
Subsequently, in Section \ref{sention:jssanet}, the proposed JSSAnet is introduced in detail.
Simulation results are provided in Section \ref{section:simulation}.
Finally, Section \ref{section:conclusion} concludes this paper.

\textit{Notation: }
We use the following notations throughout the paper.
$\mathbf{A}$ is a matrix; 
$\mathbf{a}$ is a vector; 
$a$ is a scalar;
the superscripts $(\cdot)^{*}$, $(\cdot)^{T}$, $(\cdot)^{H}$ and $(\cdot)^{-1}$ stand for  conjugate operator, transpose operator, conjugate transpose operator and matrix inverse, respectively;
$\|\mathbf{A} \|_{F} $ and $\mathrm{tr}(\mathbf{A})$ are the Frobenius norm and trace of $\mathbf{A} $, respectively;
$\mathbf{I}_{N}$ is the $N \times N$ identity matrix;
$[\mathbf{a}]_{i}$ and $[\mathbf{A}]_{i,j}$ denote $i$-th entry of $\mathbf{a}$ and entry at the $i$-th row and $j$-th column of $\mathbf{A}$;
$\mathbf{x}\sim \mathcal{CN}(\mathbf{a}, \mathbf{A})$ is a complex Gaussian vector with mean $\mathbf{a}$ and covariance matrix $\mathbf{A}$.
$\mathbb{E}\{\cdot \}$ is used to denote expectation;
$ \lceil \cdot\rceil$ and $ \lfloor \cdot\rfloor$ are ceiling and floor operations, respectively;
$\otimes$ and $\odot$ denote the Kronecker product and element-wise product, respectively;
$C(X) = \int_{0}^{X} \cos (\frac{\pi x^{2}}{2}) \mathrm{d}x $ and $S(X) = \int_{0}^{X} \sin (\frac{\pi x^{2}}{2}) \mathrm{d}x$ are Fresnel functions.

\section{System Model and Problem Formulation}\label{section:system model}
\subsection{System Model}

\begin{figure}[t]
	\centering
	\includegraphics[width=0.45\textwidth]{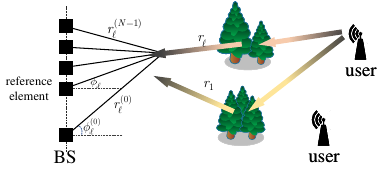}
	\caption{The near-field uplink channel with spherical-wave and  with several scatters.}
	\label{fig:channel model}
\end{figure}

We consider an uplink time division duplexing (TDD) narrowband OFDM system with $K$ subcarriers, as shown in Fig. \ref{fig:channel model}.
The base station (BS) with $N_\text{BS}$-antenna uniform linear array (ULA) and $N_\text{RF}$ RF chains is assumed to communicate with $U$ single-antenna users, where the antenna spacing $d=\frac{\lambda}{2}$ is the half of signal wavelength.
The pilot sequences emitted by users to the BS and the channel fading remain constants during the uplink channel estimation period.
In this paper, we adopt the orthogonal pilot strategy such that the channel estimation for each user is independent and a specific user is selected \cite{NearCE_Dai,Gao2019_Wideband}.
Without loss of generality, one can choose the pilot signal $s_{t,k} = 1$ in the $t$-th time slot on subcarrier $k$. 

We collect the received pilots during $T$ instants into a vector $\mathbf{y}_k\in \mathbb{C}^{TN_{\text{RF}} \times 1}$ on subcarrier $k$ and stacking these from all subcarriers into a matrix, we have obtain the overall measurement matrix $\mathbf{Y} = [\mathbf{y}_1, \cdots, \mathbf{y}_K ] \in \mathbb{C}^{TN_{\text{RF}} \times K}$ as \cite{NearCE_Dai,zhu_twc}
\begin{equation} \label{eq:Y}
	\mathbf{Y} = \mathbf{W}^H \mathbf{H} + \mathbf{V},
\end{equation}
where $\mathbf{W}=[\mathbf{W}_1, \cdots, \mathbf{W}_{\text{T}}] \in \mathbb{C}^{N\times TN_{\text{RF}}} $ is a combining matrix composed by the receive combining matrix $\mathbf{W}_t$ of size $N_\text{BS}\times N_{\text{RF}}$ for $t\in\{1, \cdots, T\}$
and $\mathbf{V} = [\mathbf{v}_1, \cdots, \mathbf{v}_K]$ of size $TN_{\text{RF}}\times K $ is defined as the effective noise matrix, where $\mathbf{v}_k = \mathbf{W}^H\mathbf{n}_k$ of size $TN_{\text{RF}}\times 1 $ and $\mathbf{n}_k \sim \mathcal{CN}(0, \sigma^2 \mathbf{I}_{N_\text{BS}})$ is the noise vector with the noise power $\sigma^2$.
Note that the entries of $\mathbf{W}$ are chosen randomly from $\frac{1}{\sqrt{N_\text{BS}}}\{-1,+1\}$.
Besides, $\mathbf{H}=[\mathbf{h}_1, \cdots, \mathbf{h}_K] \in \mathbb{C}^{N_\text{BS}\times K}$ denotes a multi-subcarrier channel matrix whose $k$-th column vector $\mathbf{h}_k\in \mathbb{C}^{N_\text{BS}\times 1}$ is the channel of a certain user on subcarrier $k$.

\subsection{Channel Model}
We commence with the near-field  mmWave channel in the spatial domain.
As depicted in Fig. \ref{fig:channel model}, we consider the users locate at the near-field region of BS.
Therefore, EM wave are incident on the antenna array of BS with the form of spherical wavefront.
Then, we adopt the widely used extended Saleh-Valenzuela multipath channel model in the frequency domain. 
Then, the near-field channel $\mathbf{h}_k$ for subcarrier $k$ can be presented as
\begin{equation}
	\mathbf{h}_k = \sqrt{\frac{N_\text{BS}}{L}}\sum_{\ell=1}^{L} \alpha_\ell e^{2\pi \tau_\ell f_k} \mathbf{a}(\theta_\ell, r_\ell),
\end{equation}
where $L\ll N$ is the number of resolvable paths, $\alpha_\ell$ and $\tau_\ell$ are the complex gain and the time delay of the $\ell$-th path, respectively.
The variable $r_\ell$ is the distance from a scatter to a reference element of antenna array in BS for the $\ell$-th path.
The variable $\theta_\ell$ is the sine of the physical AoA $\phi_\ell$ corresponding to $r_\ell$, i.e., $\theta_\ell = \sin \phi_\ell$.
Then, $\phi_\ell^{(n)}$ is the physical AoA of the $n$-th element where $n=0,1,\cdots, N_\text{BS}-1$.
In this paper, the $(\frac{N_\text{BS}}{2}-1)$-th element is set as the reference element of the ELAA, i.e., $\theta_\ell=\theta_\ell ^{\left ({N_\text{BS}}/{2}-1\right )}$ and $r_\ell = r_\ell^{\left ({N_\text{BS}}/{2}-1\right )}$.

Finally, $\mathbf{a}(\theta_\ell, r_\ell)$ is the normalized antenna array response vector (ARV) at the BS.
For the $N_\text{BS}$-element ULA, $\mathbf{a}(\theta_\ell, r_\ell)$ is written as
\begin{equation}\label{eq:sv_nf}
\begin{aligned}
	\mathbf{a}(\theta_\ell, r_\ell) = 
	\frac{1}{\sqrt{N_\text{BS}}}
	\left[e^{j\frac{2\pi}{\lambda}\left (r_{\ell}^{(0)}-r_{\ell}\right ) }, \cdots,
	e^{j\frac{2\pi}{\lambda}\left (r_{\ell}^{(N_\text{BS}-1)}-r_{\ell}\right ) } \right]^H ,
\end{aligned}
\end{equation}
where $r_\ell^{(n)}$ for $n\in\{0,1,\cdots,N_\text{BS}-1\}$ denotes the distance from a scatter to $n$-th antenna element for the $\ell$-th path component.
In the far-field communication, planar wavefronts lead to the linear distance difference of arrival of EM waves, which is calculated by
\begin{equation}\label{eq:dod_ff}
	r^{(n)}_\ell-r_\ell = - \Delta_n d \theta_\ell,
\end{equation}
where $\Delta_n = n-\frac{N_\text{BS}}{2}+1$ is the index interval between the $n$-th element and the reference element, i.e., $\Delta_n = -\frac{N_\text{BS}}{2}+1, -\frac{N_\text{BS}}{2}+2, \cdots, \frac{N_\text{BS}}{2}  $.

However, spherical wavefronts in the near-field communication result in the distance differences of arrival of EM waves is non-linear with respect to the antenna index $n$.
Accordingly, the distance difference of arrival for the $\ell$-th path is approximately expressed as \cite{zhu_twc}
\begin{equation}\label{eq:dod_nf}
	r^{(n)}_\ell-r_\ell \approx   \frac{1-\theta_\ell^2}{2r_\ell} d^2 \Delta_n^2 - \Delta_n d \theta_\ell.
\end{equation}
It reveals that the explicit distinction of expressions between near-field and far-field channels lies in the presence of quadratic term in the phase term.

\subsection{Near-field Channel Estimation Formulation}
As we know, \eqref{eq:Y} is a typical linear expression of pilot-based mmWave channel estimation, and various methods are rendered, including the LS algorithm, the linear minimum mean squared error (LMMSE) algorithm, the CS approach and the DL approach.

\textit{1) LS algorithm:}
The LS algorithm originated from minimizing $\left \|\mathbf{y}_k - \mathbf{W}^H \hat{\mathbf{h}}_k\right \|^2$.
Hence, the estimated channel $ \hat{\mathbf{h}}_k^{\text{LS}}$ for subcarriers $k$ by LS estimator is expressed as
\begin{equation}
	\hat{\mathbf{h}}_k^{\text{LS}} = \left (\mathbf{WW}^H\right )^{-1}\mathbf{Wy}_k.
\end{equation}
Hence, the LS-based estimated channel for all subcarriers $\hat{\mathbf{H}}_\text{LS}$ is $\hat{\mathbf{H}}_\text{LS}=[\hat{\mathbf{h}}_1^{\text{LS}},\hat{\mathbf{h}}_2^{\text{LS}} \cdots, \hat{\mathbf{h}}_K^{\text{LS}}] $.

\textit{2) LMMSE algorithm:}
The LS algorithm has low complexity but poor performance in low-SNR regime.
Therefore, employing the weighted matrix is to minimize the euclidean distance between the true channel and the LS-based estimated channel.
The estimated channel $ \hat{\mathbf{h}}_k^{\text{LMMSE}}$ for subcarrier $k$ by LMMSE estimator can be obtained by
\begin{equation}
	\hat{\mathbf{h}}_k^{\text{LMMSE}} = \mathbf{R}_{\mathbf{h}_k \hat{\mathbf{h}}_k^{\text{LS}}} \left( \mathbf{R}_{\mathbf{h}_k \mathbf{h}_k } +\sigma^2 \mathbf{I}_{N_\text{BS}} \right) ^{-1} \hat{\mathbf{h}}_k^{\text{LS}},
\end{equation}
where $\mathbf{R}_{\mathbf{h}_k \hat{\mathbf{h}}_k^{\text{LS}}}$ is the cross correlation matrix between $\mathbf{h}_k$ and $ \hat{\mathbf{h}}_k^{\text{LS}}$, and $\mathbf{R}_{\mathbf{h}_k \mathbf{h}_k }$ is the autocorrelation matrix of $\mathbf{h}_k$.
Hence, the LMMSE-based estimated channel for all subcarriers $\hat{\mathbf{H}}_\text{LMMSE}$ is $\hat{\mathbf{H}}_\text{LMMSE}=[\hat{\mathbf{h}}_1^{\text{LMMSE}},\hat{\mathbf{h}}_2^{\text{LMMSE}}, \cdots, \hat{\mathbf{h}}_K^{\text{LMMSE}}] $.

\textit{3) CS approach:}
The LS and LMMSE algorithms ignore the inherent property of the mmWave channels.
Thus, the CS approach is employed to reconstruct the mmWave channel by exploiting the sparse spatial nature from poor scatters, where channel representations are found from redundant dictionaries.
This problem can be presented as minimizing a $\ell_0$-norm function \cite{Wipf2010_Iterative} and given by
\begin{equation}
	\min \| \mathbf{ x}_k\|_0, \quad \mathrm{s.t.} \quad \mathbf{y}_k = \mathbf{W}^H \mathbf{\Phi x}_k ,
\end{equation}
where $\mathbf{\Phi} \in \mathbb{C}^{N_\text{BS}\times D}$ and $\mathbf{x}_k \in \mathbb{C}^{D\times 1}$ are the transform dictionary of size $D$ for $\mathbf{h}_k$ and a sparse vector, respectively.
Note that the CS-based schemes are highly sensitive to dictionary selection and the stopping criteria setting.
In far-field communication scenarios, the DFT matrix is chosen as the dictionary $\mathbf{\Phi}$ because its array response vector is discrete Fourier vector.
However, the involved distance information makes DFT matrix inefficient in near-field channel estimation.
Consequently, the design of near-field dictionaries has become an active area of research \cite{NearCE_Dai, zhu_twc}.

\textit{4) DL approach:}
DL-based approaches are widely applied in channel estimation.
The channel estimation procedure is typically transferred to a signal denoising task in the existing DL-based schemes \cite{Liu2022_Deep,Jiang2021_Dual,Xu2024_Deep}.
In the procedure, the pre-estimated channel from LS estimator is fed into the developed neural network which the output of is viewed as improved channel reconstruction.

The DL-based channel estimation procedure can be concluded into two phases: \textit{offline training phase} and \textit{online estimation phase}.
Let the training set $\mathcal{S}_\text{t}$ with size of $\|\mathcal{S}_\text{t}\|$ and testing set $\dot{\mathcal{S}}_\text{t}$ with size of $\|\dot{\mathcal{S}}_\text{t}\|$ denote as 
\begin{equation}\label{eq:dataset training}
	\mathcal{S}_\text{t} = \left\lbrace \left( \mathbf{H}_{\text{LS}}^{(1)}, \mathbf{H}^{(1)} \right), \cdots, \left( \mathbf{H}_{\text{LS}}^{(\|\mathcal{S}_\text{t}\|)}, \mathbf{H}^{(\|\mathcal{S}_\text{t}\|)} \right)  \right\rbrace 
\end{equation}
and 
\begin{equation}\label{eq:dataset testing}
	\dot{\mathcal{S}}_\text{t} = \left\lbrace \left( \dot{\mathbf{H}}_{\text{LS}}^{(1)}, \dot{\mathbf{H}}^{(1)} \right), \cdots, \left( \dot{\mathbf{H}}_{\text{LS}}^{(\|\dot{\mathcal{S}}_\text{t}\|)}, \dot{\mathbf{H}}^{(\|\dot{\mathcal{S}}_\text{t}\|)} \right)  \right\rbrace, 
\end{equation}
respectively, where $ \mathbf{H}_{\text{LS}}^{(i)}$ and $ \dot{\mathbf{H}}_{\text{LS}}^{(i)}$ are the $i$-th noisy channel from LS observations, and $\mathbf{H}^{(i)} $ and $\dot{\mathbf{H}}^{(i)}$ are corresponding true channels.
Note that there is no overlap between $\mathcal{S}_\text{t}$ and $\dot{\mathcal{S}}_\text{t}$.
The forward propagation process and parameters of the neural network are defined as $\mathcal{F}_{\mathbf{\Theta}}(\cdot)$ and $\mathbf{\Theta}$, respectively.

In training phase, the training set $\mathcal{S}_\text{t}$ is sent into the designed neural network and back propagation refines the parameters $\mathbf{\Theta}$ of $\mathcal{F}_{\mathbf{\Theta}}(\cdot)$ aiming to minimize the loss function.
In general, the empirical mean square error (MSE) criterion is utilized to the loss function, i.e.,
\begin{equation}\label{eq:loss}
	\mathcal{L}({\mathbf{\Theta}}) = \frac{1}{\|\mathcal{S}_\text{t}\|} \sum_{i=1}^{\|\mathcal{S}_\text{t}\|} \left\| \mathbf{H}^{(i)} -  \mathcal{F}_{\mathbf{\Theta}}\left (\mathbf{H}_{\text{LS}}^{(i)}\right ) \right\| _F^2.
\end{equation}
Consequently, it is achieved the well-trained network with the trained parameters $\mathbf{\Theta}^*$.

In online estimation phase, the well-trained network $\mathcal{F}_{\mathbf{\Theta}^*} $ processes the testing data $\dot{\mathbf{H}}_{\text{LS}} $ sampled from $\dot{\mathcal{S}}_\text{t}$.
The estimated channel $\hat{\mathbf{H}}_\text{net} $ is obtained by
\begin{equation}
	\hat{\mathbf{H}}_\text{net} = \mathcal{F}_{\mathbf{\Theta}^*}  \left (\dot{\mathbf{H}}_{\text{LS}} \right ).
\end{equation}

In this work, we also adopt this framework to formulate the channel estimation problem as an LS estimation denoising task and the detail of our neural network will be provided in following sections.

\section{Piecewise Linear Approximation}\label{section:subchannel}

As a data-driven approach, the performance of DNN-based channel estimation strongly depends on its ability to extract discriminative features from input data.
Several strategies are explored to improve the representation capacity of neural networks for mmWave channels.
First, a self-attention mechanism is integrated into effectively model inter-element dependencies within the channel matrix to suppress noise.
Second, inspired by CS approach exploring the inherent sparsity, the mmWave channel is projected into the angular domain via DFT and the neural network extracts its sparse features to enhance denoising performance, as demonstrated in \cite{Jiang2021_Dual,Xu2024_Deep}.

Inspired by these strategies, we propose a hybrid strategy for the near-field channel estimation, in which the sparse angular representations are embedded into the attention procedure.
However, existing transform dictionary inevitably expand the size of sparse representation matrix for XL-MIMO communications, which substantially raises the computational cost of neural networks.
Meanwhile, \eqref{eq:dod_nf} indicates that the time differences of arrival exhibit a nonlinear property with respect to the antenna index.
The non-negligible angular energy spread significantly degrades the efficiency of DFT in sparsely representing near-field channels\cite{zhu_twc}.

This section presents an approximate representation of the near-field ARV using a piecewise linear vector compatible with the DFT.
Then, a theoretical lower bound on the piecewise number is derived to satisfy the approximation criterion.
Furthermore, a theoretical upper bound is established to ensure that angular diversity across subchannels is maintained in the angular-domain projection.

\subsection{The Piecewise Linear Approximation of the Near-field ARV}

Given the non-linearity of near-field XL-MIMO channels, it is an alternative approach that the near-field ARV is approximately decomposed into multiple linear vectors catering to the linearity of DFT.
In this work, the ELAA of BS is partitioned uniformly into $M$ subarrays of size $N$, where $M = \lceil \frac{N_\text{BS}}{N} \rceil $.
Then, the antenna index set is $\mathcal{S}_i= \{(i-1)N, \cdots, iN-1\}$ for  $i\in \{1,2,\cdots, M\}$.
Accordingly, we choose the $n_i$-th element as the reference element corresponding to the $i$-th subarray, where $n_i = iN-\frac{N}{2}-1 $.

Here, we focus on a certain path component.
For the sake of notation, omit the subscript $\ell$ and define $\theta_0$ and $r_0$ as reference AoA and distance for the entire array.
Recall \eqref{eq:sv_nf} and \eqref{eq:dod_nf}, the phase difference of arrival is defined as $p_n$ corresponding to $n$-th antenna element, given by
\begin{equation}
	p_n = -\frac{2}{\lambda}\left (r^{(n)}-r_0\right )\approx  -\frac{1-\theta_0^2}{2r_0} d \Delta_n^2 + \Delta_n \theta_0,
\end{equation}
and the derivation with respect to the index interval $\Delta_n$ is 
\begin{equation}
	p_n' = -\frac{1-\theta_0^2}{r_0} d \Delta_n +  \theta_0.
\end{equation}

Then, expanding with first-order Taylor series at the reference element of the $i$-th subarray $n = n_i$, we have
\begin{equation}\label{eq:p_taylor}
	p_n = p_{n_i} +p'_{n_i} (n-n_i) + \mathcal{O}(n).
\end{equation}

\begin{figure}[t]
	\centering
	\includegraphics[width=0.45\textwidth]{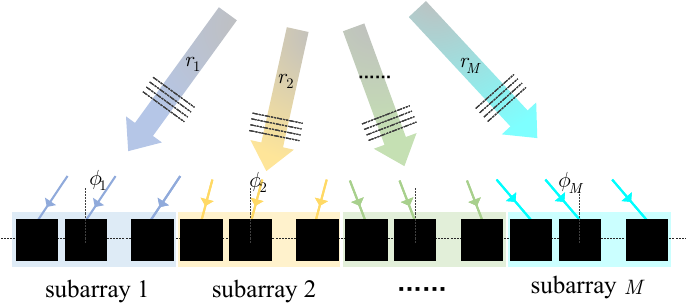}
	\caption{
		The ELAA is partitioned uniformly into $M$ subarrays.
		The EM waves impinge on the subarray with planar wavefront, while the wavefronts between subarrays exhibit spherical waves.}
	\label{fig:subantenna}
\end{figure}

Suppose the scatter locates at $(x,y)$ and the coordinate of $n$-th element is $(0,\Delta_nd)$, the spatial AoA $\theta^{(n)}$ and distance $r^{(n)}$ between the scatter and $n$-th element are computed by
\begin{equation}
	\begin{cases}
		\theta^{(n)} = \dfrac{y-\Delta_nd}{r^{(n)}}, \\
		r^{(n)} = \sqrt{x^2+(y-\Delta_nd)^2}.
	\end{cases}
\end{equation}
Then, expanding with first-order Taylor series at the reference element,i.e., $n=\frac{N_\text{BS}}{2}-1$, we have
\begin{equation}\label{eq:theta_taylor}
	\theta^{(n)} = \theta_0 -\frac{1-\theta_0^2}{r_0}d \Delta_n +\mathcal{O}(n) \approx p_n'.
\end{equation}
Substituting \eqref{eq:theta_taylor} into \eqref{eq:p_taylor}, the approximate phase vector $\mathbf{p}_i$ of the near-field ARV corresponding to $i$-th subarray can be linearly written as
\begin{equation}\label{eq:phase_sub}
	\mathbf{p}_i \approx \left [p_{n_i} + m\theta^{(n_i)}\right ]_{m=-\frac{N}{2}+1}^{\frac{N}{2}},
\end{equation}
where $ m \triangleq n-n_i \in \left\lbrace -\frac{N}{2}+1, -\frac{N}{2}+2, \cdots, \frac{N}{2}\right\rbrace  $.
It indicates that the exponential term of the subarray response vector can approximately be composed by two parts: time difference of arrival $p_{n_i}$ and spatial AoA $\theta^{(n_i)}$ corresponding to the reference element of the subarray.

For clarity, we define $ \tilde{p}_i \triangleq p_{n_i} $, $\tilde{r}_i \triangleq r^{(n_i)}$ and $\tilde{\theta}_i \triangleq \theta^{(n_i)}$.
Subsequently, according to \eqref{eq:phase_sub}, the array response vector of the $i$-th subarray is approximately rewritten as  
\begin{equation}\label{eq:sv_sub}
	[\mathbf{a}(\theta_0, r_0)]_{\mathcal{S}_i} \approx \sqrt{\frac{N}{N_\text{BS}}}e^{j\pi \tilde{p}_i}\mathbf{b}(\tilde{\theta}_i),
\end{equation}
where $	\mathbf{b}(\tilde{\theta}_i)= \frac{1}{\sqrt{N}} \left [1, e^{j\pi \tilde{\theta}_i}, \cdots, e^{j\pi (N-1)\tilde{\theta}_i} \right ]^T \in \mathbb{C}^{N\times 1}$ is a Fourier vector.
It is observed that the subarray response vector is consisted of a free-space phase delay factor of subarray $i$ and a linear Fourier vector with size of $N\times 1$.
Thus, collecting linear subarray response vectors, the array response vector $\mathbf{a}(\theta_0, r_0)$ can be piecewise represented by
\begin{equation}\label{eq:sv_linear}
	\mathbf{b}(\bm{\theta})=\sqrt{\frac{N}{N_\text{BS}}}\left [
	e^{j\pi \tilde{p}_1} \mathbf{b}(\tilde{\theta}_1)^T,\cdots,e^{j\pi \tilde{p}_M} \mathbf{b}(\tilde{\theta}_M)^T
	\right ]^T,
\end{equation}
where ${\bm{\theta}}=[\tilde{\theta}_1, \tilde{\theta}_2, \cdots, \tilde{\theta}_M]$.

Therefore, the near-field ARV can be approximated as a piecewise Fourier vector.
It is consistent with intuition that the case of a user locating at the near-field region of the ELAA can be equivalent to that of a user locating at the far-field regions of partitioned subarrays.
As shown in Fig. \ref{fig:subantenna}, the entire antenna array is divided uniformly into $M$ subarrays.
It is assumed that EM waves impinging on each subarray can be modeled as planar, while the wavefronts between subarrays exhibit spherical characteristics.
Essentially, the piecewise Fourier vector implicitly encodes the nonlinear spatial information of near-field ARV utilizing the different AoAs across subarrays.

\begin{figure}[t]
	\centering
	\includegraphics[width=0.35\textwidth]{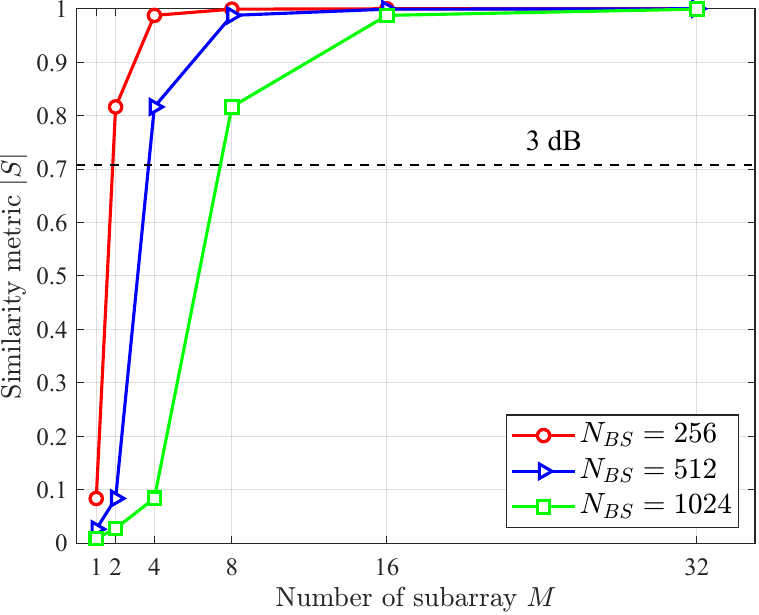}
	\caption{
		The similarity metric between $\mathbf{b}(\bm{\theta})$ and $\mathbf{a}(\theta_0, r_0)$.
		The carrier frequency $f_c$, the spatial AoA $\theta_0$ and distance $r_0$ are $60$ GHz, $\sin(\pi/12)$ and $15$ meters, respectively.
	}
	\label{fig:similarity}
\end{figure}

\subsection{Discussion on Number of Subarrays}\label{subsection:number}

Intuitively, the number of partitions determines the fidelity of the piecewise Fourier vector in approximating the near-field ARV $\mathbf{a}(\theta_0, r_0)$.
From a theoretical perspective, an insufficient number makes it difficult for the piecewise Fourier vector to precisely approximate the near-field ARV.

\begin{theorem}\label{pos:M_min}
	If the piecewise Fourier vector can approximately present the near-field ARV, the partitioned number $M$ must satisfy 
	\begin{equation}
		M \geq \left \lceil \sqrt{\frac{d}{ r_\text{min}} } \frac{N_\text{BS}}{4} \right \rceil,
	\end{equation}
	where $r_\text{min} $ is the minimum feasible distance in communication system.
\end{theorem}

\textit{~Proof:} See Appendix \ref{app:1}. \hfill $\blacksquare$

According to \textit{Theorem \ref{pos:M_min}}, for a $512$-ULA, the partitioned number must satisfy the condition of $M \geq \lceil 2.86\rceil=3$.
For implementation convenience, $M$ is set to be a power of $2$, i.e., $M=4$.
As shown in Fig. \ref{fig:similarity}, we present the similarities between  $\mathbf{b}(\bm{\theta})$ and $\mathbf{a}(\theta_0, r_0)$ corresponding to various array sizes.
It can be observed that for ULAs with $N_\text{BS}\in \{256,512, 1024\}$, the corresponding minimum values of $M$ are 2, 4 and 8, respectively, which validate the above theorem of $M \geq \lceil1.43\rceil $, $M \geq \lceil2.86\rceil$ and $M \geq \lceil5.72\rceil$.

While partitioning the ELAA facilitates the transformation of the channel into the angular domain,  it inevitably reduces the spatial angular resolution of each receive subarray, as resolution is proportional to the array aperture. 
Therefore, the degradation in angular resolution imposes a practical upper limit on the partitioned number that can be employed.

To elaborate, as shown in Fig. \ref{fig:subantenna}, the near-field channel can be stacked by $M$ subchannels, i.e.,
\begin{equation}
	\mathbf{H}=\left [ \mathbf{H}_1^T, \mathbf{H}_2^T, \cdots, \mathbf{H}_M^T  \right ]^T,
\end{equation}
where $\mathbf{H}_i=[\mathbf{h}_1^{(i)}, \cdots, \mathbf{h}_K^{(i)}] \in \mathbb{C}^{N\times K}$ is the $i$-th subchannel between the user and subarray $i$  for all subcarriers and $\mathbf{h}_k^{(i)}$ is the $i$-th subchannel at subcarrier $k$.
Accordingly, the angular-domain piecewise representation of $\mathbf{H}$ is determined by the subchannels $\mathbf{H}_i$, i.e.,
\begin{equation}\label{eq:dft H}
	\begin{aligned}
		\mathcal{D}(\mathbf{H}) &= (\mathbf{I}_M \otimes \mathbf{\Phi}) ^H \mathbf{H}
		\\
		&=\left [ (\mathbf{\Phi} ^H \mathbf{H}_1)^T, (\mathbf{\Phi} ^H \mathbf{H}_2)^T, \cdots, (\mathbf{\Phi} ^H \mathbf{H}_M)^T  \right ]^T,
	\end{aligned}
\end{equation}
where $\mathbf{\Phi} = [\mathbf{b}(\varphi_1), \mathbf{b}(\varphi_2), \cdots, \mathbf{b}(\varphi_N)]$ is a unitary DFT matrix with size of $N\times N$ and $\varphi_n = \frac{1}{N}(n-\frac{N+1}{2})$ for $n=1, 2\cdots, N$.
Accordingly, the beam pattern vector $\mathbf{c}_{k,\ell}^{(i)}$ of the $\ell$-th path component corresponding to $\mathbf{h}_k^{(i)}$ in angular domain is determined as 
\begin{equation}
	\begin{aligned}
		\mathbf{c}_{k,\ell}^{(i)} 
		&= \mathbf{\Phi}^H \mathbf{b}(\tilde{\theta}_{\ell,i})
		\\
		&= [\Xi(\tilde{\theta}_{\ell,i}-\varphi_1), \Xi(\tilde{\theta}_{\ell,i}-\varphi_2), \cdots, \Xi(\tilde{\theta}_{\ell,i}-\varphi_N)]^T,
	\end{aligned}
\end{equation}
where $\tilde{\theta}_{\ell,i} $ is the spatial AoA of the $\ell$-th path for subarray $i$ and $\Xi(x) = \frac{\sin N\pi x}{N\sin \pi x}$ is the Dirichlet sinc function, whose power is concentrated near $x=0$.
Thus, $|[\mathbf{c}_{k,\ell}^{(i)}]_n|^2 $ presents the power of the $n$-th entry of $\mathbf{c}_{k,\ell}^{(i)}$.

\begin{figure}[t]
	\centering
	\begin{subfigure}[t]{0.8\linewidth}
		\centering	
		\includegraphics[width=\textwidth]{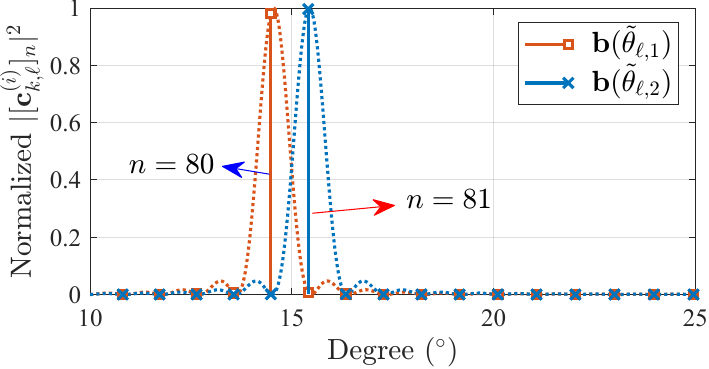}
		\caption{$M=2$}
		\label{fig:dft_sub2}
	\end{subfigure}
	  
	\begin{subfigure}[b]{0.8\linewidth}
		\centering
		\includegraphics[width=\textwidth]{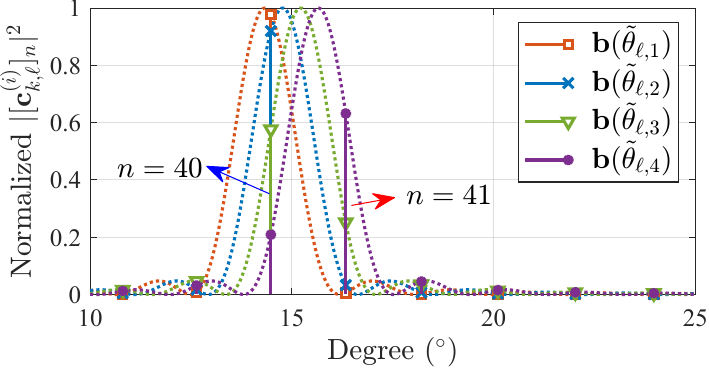}
		\caption{$M=4$}
		\label{fig:dft_sub4}
	\end{subfigure}
	\caption{
		The normalized power distributions of $\mathbf{c}_{k,\ell}$ for the $\ell$-th path component at $f_k=60$ GHz, where $N_\text{BS}=256$, $\phi_\ell=15 ^\circ$ and $r_\ell=20$ meters.
		(a) The case of $M=2$; (b) The case of $M=4$.
	} 
	\label{fig:beam patterns}
\end{figure}

It has been observed that the adverse effects of near-field conditions on channel estimation become increasingly pronounced as users move closer to the BS \cite{zhu_twc,Wan2024_Field}.
To guarantee the lower bound of near-field channel estimation, it is required that beam patterns of subchannels represented by different Fourier vectors of $\mathbf{\Phi}$ within the region of the minimum feasible propagation distance, 
i.e., $\arg\max_n |[\mathbf{c}_{k,\ell}^{(i)}]_n|^2 \neq \arg\max_n |[\mathbf{c}_{k,\ell}^{(j)}]_n|^2$ for $i\neq j$ when $r=r_\text{min}$.
However, as the subarray aperture decreases below a certain threshold, it inevitably leads to cases where at least two subchannels exhibit a nearly identical angular distribution.
For example, a near-field scenario with $N_\text{BS}=256$, $\phi_\ell = 15 ^\circ$ and $f_c=60$ GHz is considered, the power distributions of beam patterns of the $\ell$-th path component are shown in Fig. \ref{fig:beam patterns}.
We observe that when the ELAA is partitioned into $2$ subarrays, each subchannel exhibits a dominant beam pattern with a distinct index, namely, $[\mathbf{c}_{k,\ell}^{(1)}]_{80}$ and $[\mathbf{c}_{k,\ell}^{(2)}]_{81}$.
However, when the ELAA is partitioned into $4$ subarrays, subchannel $1$ and $2$ share the same dominant beam, namely, $ \mathbf{b}(\varphi_{40})$ and $[\mathbf{c}_{k,\ell}^{(1)}]_{40}\approx [\mathbf{c}_{k,\ell}^{(2)}]_{40}$.
Therefore, $\mathcal{D}(\mathbf{H}_1)$ and $\mathcal{D}(\mathbf{H}_2)$ share almost common spatial angular features, making it difficult for the neural network to extract the specific spatial features of different subchannels.
Accordingly, we provide the upper bound on the number of subchannels below.

\begin{theorem}\label{pos:M_max}
	If the number of partitioned subarrays $M$ only satisfies
	\begin{equation}
		M \leq \left \lfloor \sqrt{\frac{(1-\theta_\text{sec}^2)d}{2 r_\text{min}}  } N_\text{BS}\right \rfloor,
	\end{equation}
	an DFT matrix with size of $N\times N$ preserves diverse angular information across subchannels when channel near-field effect with $r=r_\text{min}$ is most pronounced.
\end{theorem}

\textit{~Proof:} See Appendix \ref{app:2}. \hfill $\blacksquare$

In this work, we propose to decompose the near-field channel into multiple subchannels and embed their joint angular representations into the attention-based channel estimation framework.
Meanwhile, \textit{Theorem \ref{pos:M_min}} and \textit{Theorem \ref{pos:M_max}} provides the theoretical bounds for the number of partitioned subchannels for the design of our neural network.

\section{Joint Subchannel-Spatial-Attention Network for Near-Field Channel Estimation}\label{sention:jssanet}
In this section, we propose a joint subchannel-spatial-attention network (JSSAnet) for near-field channel estimation based on the ELAA partitioning strategy.
The network provides mutually independent and personalized attention to the distinct spatial information of each subchannel in parallel, while also fusing inter-subchannel information to model the nonlinear representation for near-field channels.
\begin{figure*}[t]
	\centering
	\begin{subfigure}[t]{0.8\linewidth}
		\centering	
		\includegraphics[width=\textwidth]{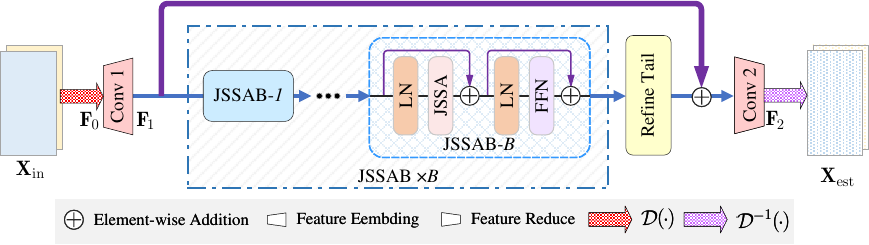}
		\caption{The structure of the proposed JSSAnet for near-field channel estimation.}
		\label{fig:network}
		\vspace{0.3cm}
	\end{subfigure}
	\begin{subfigure}[b]{0.48\linewidth}
		\centering
		\includegraphics[width=\textwidth]{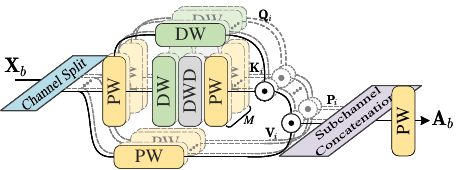}
		\caption{Joint Subchannel-Spatial-Attention (JSSA) Module}
		\label{fig:sa}
	\end{subfigure}
	\begin{subfigure}[b]{0.28\linewidth}
		\centering
		\begin{subfigure}[b]{0.95\linewidth}
			\centering	
			\includegraphics[width=\textwidth]{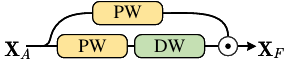}
			\caption{Forward-feed network (FFN) }
			\label{fig:ffn}
			\vspace{0.2cm}
		\end{subfigure}	
		\begin{subfigure}[b]{0.95\linewidth}
			\centering
			\includegraphics[width=0.95\textwidth]{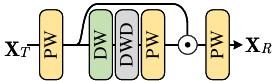}
			\caption{Refine Tail}
			\label{fig:tail}
		\end{subfigure}
	\end{subfigure}
	
	\caption{
		Decomposition of the proposed JSSAnet: Overall Architecture and functional Modules.
	}
\end{figure*}

\subsection{The Structure Design of JSSAnet}

As shown in Fig. \ref{fig:network}, the proposed JSSAnet applies a standard autoregressive architecture with a residual shortcut connection, integrating multiple functional modules to implement near-field channel estimation.

In detail, a DFT operation $\mathcal{D}(\cdot)$ and an inverse DFT (IDFT) operation $\mathcal{D}^{-1}(\cdot)$ provide bi-directional mapping between the channel matrix and its angular-domain representation.
The convolutional layers Conv 1 and Conv 2, both with $\{3, 1\}$\footnote{We denote $\{a, b\}$ as the convolution layer, where $a$ is a $a\times a$ filter, $b$ is a stride of $b$.} are utilized to embed features\footnote{We adopt `feature' as the replacement term to avoid confusion with `channel' of the communication system in this paper.} to $C>2$ and reduce features back to the original input dimension before IDFT operation, respectively.
Notice that Conv 1 performs feature embedding on $M$ subchannels simultaneously and independently and features of subchannels are stacked to $\mathbf{F}_1\in \mathbb{R}^{N_\text{BS}\times K\times C}$.
$B$ consecutive joint subchannel-spatial-attention blocks (JSSABs) are main body,  enabling joint exploitation of subchannel spatial sparsity.
To further enhance feature representation, a refinement tail module is appended after the final JSSAB to fuse and refine features across all subchannels.

\subsection{DFT and IDFT Operations}
Different from traditional near-field channel estimation based on attention mechanisms, we exploit the prior information of channel sparsity in an attention procedure.
Prior studies have extensively investigated the sparse representation of near-field channels \cite{Lu2023_Field,Near_Zhang}. 
However, the expanded dictionaries, due to the additional spatial distance information, inevitably increase the size of the sparse matrix, raising the computational overhead.
Accordingly, the channel is divided into $M$ subchannels with size of $N\times K$, and each subchannel is projected into the angular domain using an DFT matrix $\mathbf{\Phi}$ with size of $N\times N$, following a piecewise Fourier vector representation.

Based on \textit{Theorem \ref{pos:M_min}} and \textit{Theorem \ref{pos:M_max}}, the channel is separated uniformly into $M$ subchannels and each subchannel is converted into an angular representation matrix by $\mathbf{\Phi}$ before Conv 1.
The output $\mathbf{F}_2 \in \mathbb{R}^{N_\text{BS}\times K \times 2}$ of Conv 2 is converted to the estimated channel matrix through IDFT operation.
Since the neural network processes real-valued data, the real and imaginary parts of the complex channel matrix are decomposed and stacked into a 2-feature tensor.
Then, the input and corresponding true channel labels of datasets \eqref{eq:dataset training} and \eqref{eq:dataset testing} are processed as
\begin{align}
	&\mathbf{X}_{\text{in}}\doteq \left\lbrace \Re(\mathbf{H}_{\text{LS}}), \Im(\mathbf{H}_{\text{LS}})\right\rbrace \in \mathbb{R}^{N_\text{BS}\times K \times 2} ,\\
	&\mathbf{X}_{\text{gt}}\doteq \left\lbrace \Re({\mathbf{H}}), \Im({\mathbf{H}})\right\rbrace \in \mathbb{R}^{N_\text{BS}\times K  \times 2}.
\end{align}
Then, \eqref{eq:dft H} is rewritten as the DFT operation on tensor $\mathbf{X}_\text{in}$\footnote{Denote $\mathbf{X}_\text{in,i}$ as the $i$-th feature matrix of tensor $\mathbf{X}_\text{in}$. 
e.g., $\mathbf{X}_\text{in,1}=\Re(\mathbf{H}_\text{LS})$ and $\mathbf{X}_\text{in,2}=\Im(\mathbf{H}_\text{LS})$.} 
to obtain tensor $\mathbf{F}_0$\footnote{Denote $\mathbf{F}_{j,i}$ as the $i$-th feature matrix of tensor $\mathbf{F}_j$.}, i.e., 
\begin{equation}
	\begin{aligned}
		\mathcal{D}(\mathbf{X}_\text{in}) = \begin{bmatrix}
			\Re(\tilde{\mathbf{\Phi}})^T & \Im(\tilde{\mathbf{\Phi}})^T\\
			-\Im(\tilde{\mathbf{\Phi}})^T & \Re(\tilde{\mathbf{\Phi}})^T\\
		\end{bmatrix}
		\begin{bmatrix}
			\mathbf{X}_\text{in,1}\\ 
			\mathbf{X}_\text{in,2}
		\end{bmatrix}=\begin{bmatrix}
		\mathbf{F}_{0,1}\\
		\mathbf{F}_{0,2}
		\end{bmatrix},
	\end{aligned}
\end{equation}
where $\tilde{\mathbf{\Phi}}=\mathbf{I}_M \otimes \mathbf{\Phi}$.
The angular-domain tensor corresponding to subchannel $\mathbf{H}_i$ is $[\mathbf{F}_0]_{i} \in \mathbb{R}^{N\times K\times 2}$, i.e., $ [\mathbf{F}_0]_i \triangleq [\mathbf{F}_0]_{(i-1)N:iN,:,:}$ for $i=1,2,\cdots, M$.
$[\mathbf{F}_0]_i$ for $i=1,2,\cdots, M$ are concatenated along the first dimension, we have 
\begin{equation}
	\mathbf{F}_0 \doteq \{[\mathbf{F}_0]_1;[\mathbf{F}_0]_2;\cdots; [\mathbf{F}_0]_M\}.
\end{equation}
The IDFT operation on $\mathbf{F}_2\in \mathbb{R}^{N_\text{BS}\times K\times 2}$ is written by
\begin{equation}
	\mathcal{D}^{-1}(\mathbf{F}_2)=
	\begin{bmatrix}
		\Re(\tilde{\mathbf{\Phi}}) & -\Im(\tilde{\mathbf{\Phi}})\\
		\Im(\tilde{\mathbf{\Phi}}) & \Re(\tilde{\mathbf{\Phi}})\\
	\end{bmatrix}
	\begin{bmatrix}
		\mathbf{F}_{2,1}\\
		\mathbf{F}_{2,2}
	\end{bmatrix}=\begin{bmatrix}
	\mathbf{X}_\text{est,1}\\ 
	\mathbf{X}_\text{est,2}
	\end{bmatrix},
\end{equation}
where $\mathbf{X}_\text{est} \doteq \{\Re(\hat{\mathbf{H}}), \Im(\hat{\mathbf{H}})\}\in \mathbb{R}^{N_\text{BS}\times K\times 2}$ is the estimated channel tensor by the JSSAnet, i.e., $\hat{\mathbf{H}} = \mathbf{X}_\text{est,1} + j \mathbf{X}_\text{est,2}$.

\subsection{The Implementation of JSSAB}
Cater to the ELAA partitioning, the design of JSSAB follows a decoupling-fusion strategy, that subchannels are first decoupled and processed independently in the joint subchannel-spaital-attention (JSSA) layer and the extracted features of subchannels are fused in a forward-feed network (FFN) layer, as described in Fig. \ref{fig:network}.
To preserve instance-specific details, a layer normalization (LN) layer is applied prior to both the JSSA and FFN layers.

\subsubsection{JSSA layer}
As a key layer of JSSAB, the JSSA layer provide the personalized and independent attentions for individual subchannels in parallel.
Specifically, the unique spatial features of each subchannel are emphasized independently and simultaneously through their dedicated attention implementations.
As shown in Fig. \ref{fig:sa}, the input tensor $\mathbf{X}_{b} \in \mathbb{R}^{N_\text{BS}\times K\times C}$ for $b=1,\cdots, B$ is divided to $M$ segments $\{[\mathbf{X}_b]_i \in \mathbb{R}^{N\times K\times C}\}_{i=1}^M $ along the first dimension by the channel-split operation, with each segmented tensor corresponding to one subchannel, i.e., $[\mathbf{X}_b]_i$ corresponds to the feature tensor of $\mathcal{D}(\mathbf{H}_i)$.
Subsequently, all segmented tensors $[\mathbf{X}_b]_i$ are processed in parallel through their respective attention implementations.

Here, we illustrate the attention implementation corresponding to the $i$-th subchannel in JSSAB.
As shown in Fig. \ref{fig:sa}, the query $\mathbf{Q}_i\in \mathbb{R}^{N\times K\times C}$, key $\mathbf{K}_i\in \mathbb{R}^{N\times K\times C}$, and value $\mathbf{K}_i\in \mathbb{R}^{N\times K\times C}$ for $i$-th subchannel at the $b$-th JSSAB  are constructed from $[\mathbf{X}_b]_i$ as
\begin{align}
	\mathbf{Q}_i&= f_{DW}(f_{PW}([\mathbf{X}_b]_i)),
	\\
	\mathbf{K}_i&= f_{DLKC}(f_{PW}([\mathbf{X}_b]_i)),
	\\
	\mathbf{V}_i&= f_{PW}([\mathbf{X}_b]_i),
\end{align}
where $f_{DW}(\cdot) $, $f_{PW}(\cdot) $ and $f_{DLKC}(\cdot) $ denote the depth-wise (DW) convolution, the point-wise (PW) convolution and the decomposed large-kernel convolution (DLKC), respectively.
Then, the $(\mathbf{Q}_i, \mathbf{K}_i, \mathbf{V}_i)$ projection $\mathbf{P}_i$ for subchannel $i$ can be written as
\begin{align}
	\mathbf{M}_i= \mathbf{Q}_i \odot \mathbf{K}_i, \label{eq:map}\\
	\mathbf{P}_i= \mathbf{M}_i \odot \mathbf{V}_i, \label{eq:project}
\end{align}
where $\mathbf{M}_i$ is spatial-attention map for the subchannel $i$.
Accordingly, $M$ joint attention implementations comprise our JSSA mechanism.

Equations \eqref{eq:map} and \eqref{eq:project} indicate the proposed JSSA differs from the used widely self-attention.
The JSSA mechanism adopt the element-wise product to calculate the individual attention map of the subchannel, rather than matrix multiplication.
Although the element-wise product implements a lightweight attention mechanism, it inherently suffers from limitations in modeling global contextual features compared to the matrix multiplication.
To address this issue, DLKC is introduced to enhance the attention mechanism’s capability of capturing long-range dependencies, owing to its expanded large receptive field.
In essential, the DLKC is a low complexity variant of the large-kernel convolutions (LKC) proposed in \cite{Guo2022_Visual}.
In detail, a LKC with an $a\times a $ filter is decomposed into the cascading  $(2d-1)\times (2d-1)$ depth-wise (DW) $f_{DW}(\cdot)$, $\lceil \frac{a}{d}\rceil \times \lceil \frac{a}{d}\rceil$ depth-wise dilation (DWD) $f_{DWD}(\cdot)$ with dilation $d$ and point-wise (PW) $f_{PW}(\cdot)$ convolutions.
The decomposition process is formulated as DLKC, i.e., $f_{DLKC}(\cdot) = f_{PW}(f_{DWD}(f_{DW}(\cdot)))$.

Moreover, it is clear that the projection of the proposed attention map removes the softmax function typically used in standard self-attention mechanisms, whose effectiveness in recognition tasks has been demonstrated in \cite{Guo2022_Visual}.
In total, the computational complexity of calculating JSSA maps grows linearly with the ELAA aperture, i.e., the order of $\mathcal{O}(N_\text{BS})$.

Subsequently, the projections for all subcannels are concatenated to a tensor $\mathbf{P}_b$ along the first dimension, i.e., $\mathbf{P}_b = \{\mathbf{P}_1; \mathbf{P}_2;\cdots;\mathbf{P}_M\}$
and $\mathbf{P}_b$ is sent into the PW convolution to fuse cross-subchannel spatial features and obtain $\mathbf{A}_b \in \mathbb{R}^{N_\text{BS}\times K\times C}$.

\subsubsection{FFN layer}
The FFN layer plays a crucial role in transformer networks.
Given that the global characteristics of the near-field channel are distributed among subchannels, the FFN in JSSAB aims to fuse the independently extracted spatial features from the JSSA layer, enhancing the nonlinear representation of JSSAnet.
To reduce computational overhead, we replace fully connected layers with a lightweight attention module comprising DW and PW convolutions, as illustrated in Fig. \ref{fig:ffn}.
Given the tensor $\mathbf{X}_A \in \mathbb{R}^{N_\text{BS}\times K\times C}$ from the LN layer, the flow procedure of FFN is represented as
\begin{equation}
	\mathbf{X}_F = f_{PW}(\mathbf{X}_A) \odot f_{DW}(f_{PW}(\mathbf{X}_A)).
\end{equation}

\subsection{Implementation of Refine Tail}
We employ the refine tail module in the tail of the autoregressive attention backbone.
In detail, the large-kernel attention (LKA) module proposed in \cite{Guo2022_Visual} with a DLKC is enclosed between two PW convolutions as shown in Fig. \ref{fig:tail}.

The refine tail module emphasizes the refinement of spatial representations across different subchannels. 
On the one hand, all subchannel features with size of $N\times K\times C$ are concatenated into a composite image for processing by the LKA, which enhances inter-subchannel dependencies.
On the other hand, it facilitates fine-grained information extraction tailored to individual subchannels.

\section{Numerical Results}\label{section:simulation}
In this section, we present evaluations of the proposed JSSAnet for near-field channel estimation.

\begin{table}[t] 
	\centering
	\caption{The Simulation Dataset Configurations}
	\label{table:datasetConfig}
	\begin{tabular}{c c}
		\toprule[0.3mm]
		\textbf{Parameters} & \textbf{Value}\\
		\toprule[0.2mm]
		The number of BS antennas $N_\text{BS}$ & $256$ \\
		Carrier frequency $f_c$ & $60$ GHz \\
		Bandwidth $f_B$ & $100$ MHz\\
		The number of subcarriers $K$ & $32$\\
		The number of paths $L$ & $L\sim Poisson(6)$\\
		The distribution of angle $\phi$ & $\phi \sim \mathcal{U}(-\frac{\pi}{3}, \frac{\pi}{3}) $\\
		The Rayleigh distance $d_R$ & $163.84$ m \\
		The distance between BS and user  $r$ & $r\sim \mathcal{U}(5, d_R)$ m\\
		Channel model realization & Extended Saleh-Valenzuela\\
		Signal-to-noise ratio SNR (dB)& $\{-10,-5,0,5,10\}$\\
		Dataset size & $2\times 10^4$\\
		Dataset split (train:test) & $4:1$\\
		\toprule[0.3mm]
	\end{tabular}
\end{table}
\begin{table}[t] 
	\centering
	\caption{Training Settings of the JSSAnet}
	\label{table:networkParam}
	\begin{tabular}{c c}
		\toprule[0.3mm]
		\textbf{Parameters} & \textbf{Value}\\
		\toprule[0.2mm]
		Embedding features $C$ & $20$\\
		Conv1, Conv2 & $\{3,1\} $ \\
		DLKC (DW-DWD-PW) & $35: 7\text{-}9(4)\text{-}1$  \\
		DW of the FFN & $\{7,1\}$\\
		The partitioned number $ M$ & $2$\\
		JSSA blocks $B$ & $3$ \\
		Batch size & $32$\\
		Training epoch $S$ & $100$\\
		Learning rate $\gamma_t$ & $\gamma_0=1\times 10^{-3}$ with warmup\\
		Scheduler & Cosine annealing strategy\\
		Training optimizer& Adam optimizer\\
		Weight decay& $1\times 10^{-4}$\\
		\toprule[0.3mm]
	\end{tabular}
\end{table}

\subsection{Simulation Setups}
The dataset comprising a total of $2\times 10^{4}$ near-field channel realizations is generated by applying extended Saleh-Valenzuela model based on the detailed environment parameters summarized in Table \ref{table:datasetConfig}.
Notice that the distance between the BS and a user are uniformly chosen from $[5, d_R]$, the number of resolvable paths follows Poisson distribution with mean $6$, and the corresponding physical AoA $\phi$ are chosen uniformly from $[-\frac{\pi}{3},\frac{\pi}{3}]$.
Besides, we employ a $4:1$ train-test split of the total channel dataset, where the testing set is excluded from offline training and used exclusively to evaluate network performance during online channel estimation.

Furthermore, the detailed training settings of the JSSAnet are all based on the Table \ref{table:networkParam}.
The involved convolutions Conv 1 and Conv 2 are implemented by a $3\times 3$ filter with stride $1$, and DW convolutions of the FFN is employed a $7\times 7$ filter with stride $1$.
In our JSSAnet, we employ the $35\times 35$ LKC, which is decomposed by the cascading $7\times 7$ DW, $9\times 9$ DWD with dilation $4$ and PW convolutions, denoted as `$35: 7\text{-}9(4)\text{-}1$'.
Based on \textit{Theorem \ref{pos:M_min}} and \textit{Theorem \ref{pos:M_max}}, we have $\lceil1.43\rceil \leq M \leq \lfloor 2.02 \rfloor$, i.e., the partitioned number is set as $M=2$.
Then, the learning rate $\gamma_s$ is set increase linearly in the first 5 epochs and update based on cosine annealing strategy, i.e., 
\begin{equation}
	\gamma_s= 
	\begin{cases}
		\dfrac{\gamma_0}{6-s}, & 1 \leq s \leq 5, \\ 
		\dfrac{\gamma_0}{2}\left(1+\cos \left(\dfrac{s-5 }{S-4}\pi\right)\right), & 6 \leq s \leq S.
	\end{cases}
\end{equation}

The normalized mean-squared error (NMSE) is selected as the performance metric, which is defined as $\mathrm{NMSE} = \mathbb{E}\left\lbrace 
\frac{\lVert \mathbf{H}- \hat{\mathbf{H}}  \rVert_{F}^{2} }{\lVert \mathbf{H} \rVert_{F}^{2}} 
\right\rbrace$,
where $\mathbf{H}$ and $\hat{\mathbf{H}}$ are the true channel and the estimated channel, respectively.
Then, we evaluate performance of the proposed JSSAnet for near-field channel estimation compared to other schemes:
\begin{itemize}
	\item LS and LMMSE algorithms: We set $TN_\text{RF}= 256$.
	\item Classical CNN \cite{Dong2019_CNN_CE}: This scheme applies $B=5$ convolution layers with the $3\times 3$ filters and the first convolution layer expands $C=48$ features.
	\item DuCNN \cite{Jiang2021_Dual,Xu2024_Deep}: This scheme also considers the channel spatial sparsity and employs the residual mapping of double CNNs.
	Two CNNs with $B=5$ convolution layers with the $3\times 3$ filters and $C=48$ are employed with $\mathbf{H}_\text{LS}$ and $\mathbf{\Phi}^H \mathbf{H}_\text{LS}$ as their respective inputs, where $\mathbf{\Phi}$ is the DFT matrix with size of $N_\text{BS}\times N_\text{BS}$.
	\item Simultaneous OMP (SOMP) with polar-based dictionary with size of $N_\text{BS}\times D$, where $D_s>N_\text{BS}$ \cite{NearCE_Dai}, for the near-field channel estimation.
	\item Spatial attention network (SAN) \cite{zhu_tcom}: Since the common self-attention computational cost suffers from the quadratic complexity in \cite{Fan2024_Spatial,Luan2023_Channelformer}, the huge size of near-field channel compounds this obstacle.
	Alternatively, we employ a transformer variant \cite{zhu_tcom} that applies spatial-attention with linear complexity, where $C=20$ embedding features, $B=4$ attention blocks and $h=2$ heads are employed.
	\item Joint subchannel-attention network (JSAnet): This scheme removes DFT and IDFT operations without considering spatial angular nature of near-field channels.
	JSAnet is considered as the ablation experiment to verify the effectiveness of our JSSAnet.
\end{itemize}
Notice that the neural network model in the final epoch is selected as the evaluated model.

\begin{figure}[t]
	\centering	
	\includegraphics[width=0.45\textwidth]{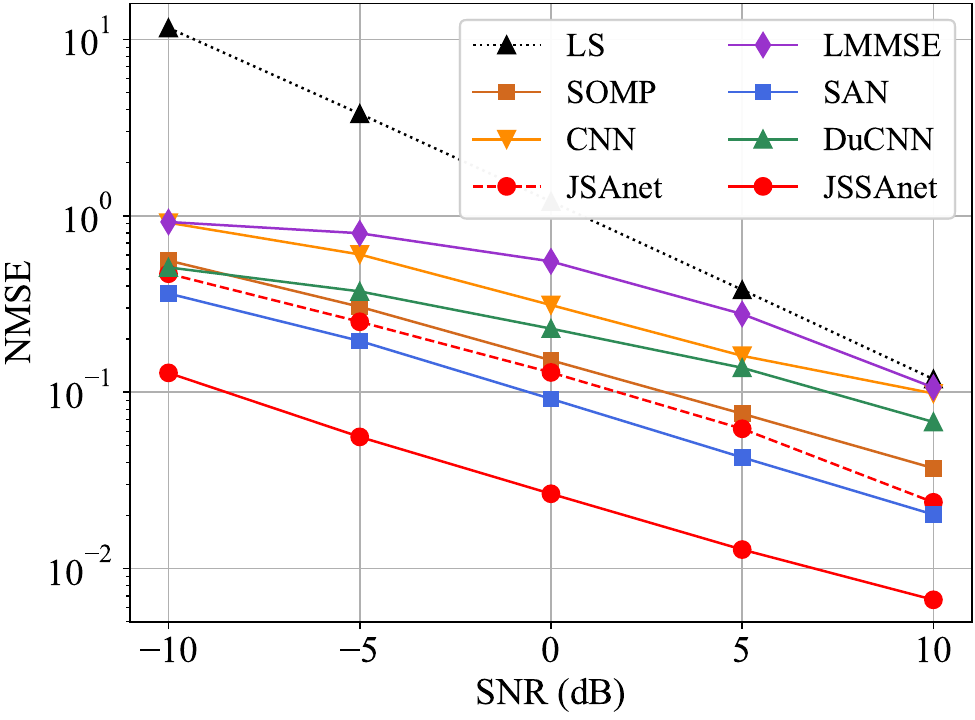}
	\caption{NMSE performance comparison of different schemes in different SNRs, where $r \sim \mathcal{U}(10, 20) $ m and $L=6$. } 
	\label{fig:snr_nmse}
\end{figure}

\subsection{Performance Comparison with Existing Methods}

Fig. \ref{fig:snr_nmse} illustrates the NMSE performance across various SNRs, where the distances are chosen from $\mathcal{U}(10, 20) $ meters and the number of paths is fixed as $L=6$.
Attention-based schemes for near-field channel estimation significantly outperform other methods.
It is clear that the proposed JSSAnet achieves the best excellent NMSE performance.
LS and LMMSE estimators suffer from the poor performance, particularly in the low SNR regime.
This is due to LS being a linear estimator with limited denoising capability.
LMMSE leverages prior channel statistics to compute a linear weighting matrix that reduces the error of $\hat{\mathbf{H}}_\text{LS}$, but it fails to exploit the inherent spatial sparsity in near-field channels.
It is observed that both CNN and DuCNN perform slightly better than LMMSE.
Due to the spatial nonstationarity of near-filed channels \cite{zhu_tcom}, both models struggle to effectively capture the complex spatial features with fixed receptive fields.
Despite DuCNN introduces the angular representation of the near-field channel via DFT for denoising, the angular spread renders the DFT inefficient\cite{zhu_twc}, limiting DuCNN to extract spatial sparse information.
The acceptable performance of SOMP is attributed to the effective exploration of near-field channel sparsity which serves as a boundary between the attention-based and other schemes.
However, its performance is highly sensitive to the dictionary selection, 
The attention mechanism contributes exceptional performance in channel denoising for SAN, JSAnet and JSSAnet.
SAN treats the channel as an image and directly applies spatial attention, but it is hard to explore efficiently the spatial natures of near-field channels without prior sparse information.
As a result, its performance only slightly outperforms SOMP and JSAnet..
As an ablation baseline, JSAnet exhibits limitations of subchannel attention via element-wise product in capturing both local and global features compared to the attention generated by matrix multiplication.
However, the proposed JSSAnet capitalizes both intra-subchannel spatial-angular consistency and inter-subchannel spatial-angular diversity to enable DFT effective for subchannels.
Moreover, our JSSA enables focusing on learning spatial sparsity patterns, yielding superior performance in channel estimation.

\begin{figure}[t]
	\centering	
	\includegraphics[width=0.45\textwidth]{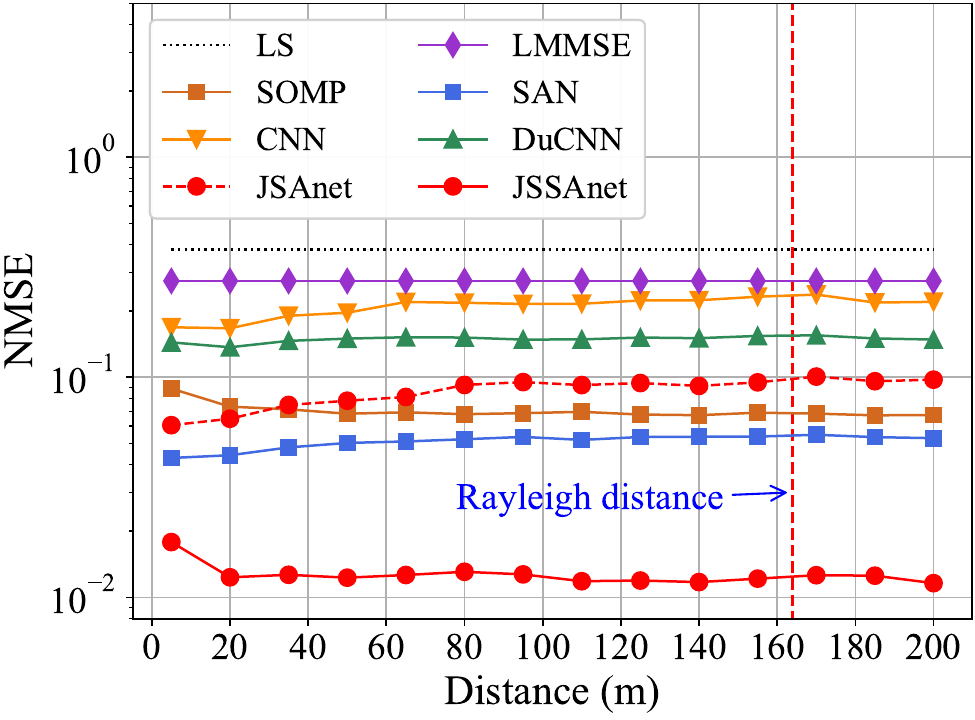}
	\caption{Comparison of the NMSE performance for distance from $5$ to $200$ meters with SNR $=5$ dB.} 
	\label{fig:dis_nmse}
\end{figure}
\begin{figure}[t]
	\centering	
	\includegraphics[width=0.45\textwidth]{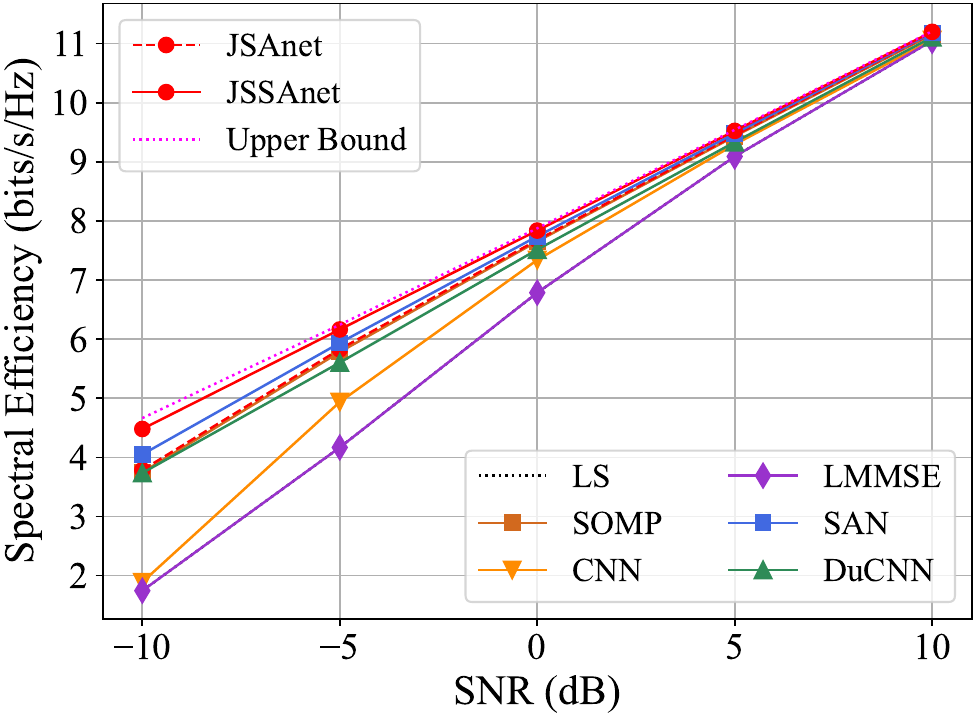}
	\caption{The spectral efficiency performance of different schemes across SNRs.	} 
	\label{fig:se_snr}
\end{figure}

We present the NMSE performance against various distance under SNR $=5$ dB, as depicted in Fig. \ref{fig:dis_nmse}.
The distance between the BS and the user ranges from 5 to 200 meters, covering both near-field and far-field regions.
It is important to note that all evaluated networks are trained solely on data of near-field region.
It can be observed that our JSSAnet significantly outperforms others and maintains stable performance in both near-field and far-field regions.
Similarly, CNN and DuCNN perform worse than other methods, except for LS and LMMSE.
Benefiting from that spatial attention excels at inter-element learning capabilities, SAN achieves the second-best NMSE performance, following our JSSAnet. 
As the distance to the BS increases, the polar-domain dictionary owns better orthogonality,, leading to improved performance of SOMP.
Conversely, the NMSE performance of JSAnet degrades at longer distances, particularly JSAnet underperforms SOMP for $r >35$m.
This is because increasing distance diminishes spherical wave effects and reduces inter-subchannel divergence, making it more challenging for JSAnet to directly extract near-field characteristics from subchannel matrices without prior spatial sparsity.
The joint subchannel spatial information makes our JSSAnet exhibit better performance and higher robustness across different distances.

In general, the spectral efficiency (SE) also serves as a performance metric.
Leveraging the channel reciprocity in TDD systems, the estimated channel is adopted to implement the downlink transmission with the MRT scheme.
The spectral efficiency is calculated by $ \mathrm{SE}=	\frac{1}{K}\sum_{k=1}^{K} \log_2\left [ 1+ \frac{|\hat{\mathbf{h}}_k^H \mathbf{h}_k|^2}{\sigma^2|\hat{\mathbf{h}}_k|^2} \right ] $.
Fig. \ref{fig:se_snr} presents the spectral efficiency across different SNRs where $r$ is chosen from $\mathcal{U}(10, 20)$ m.
Similar to NMSE performance, the SE performance of SOMP serves as lower bound of the attention-based schemes.
Then, the lower SE performances of both CNN and DuCNN reveals the weakness in near-field channel estimation tasks.
It is observed that the SE performance of our proposed JSSAnet surpasses that of other methods and closely approaches the theoretical upper bound.

\begin{figure*}[t]
	\centering
	\includegraphics[width=0.95\textwidth]{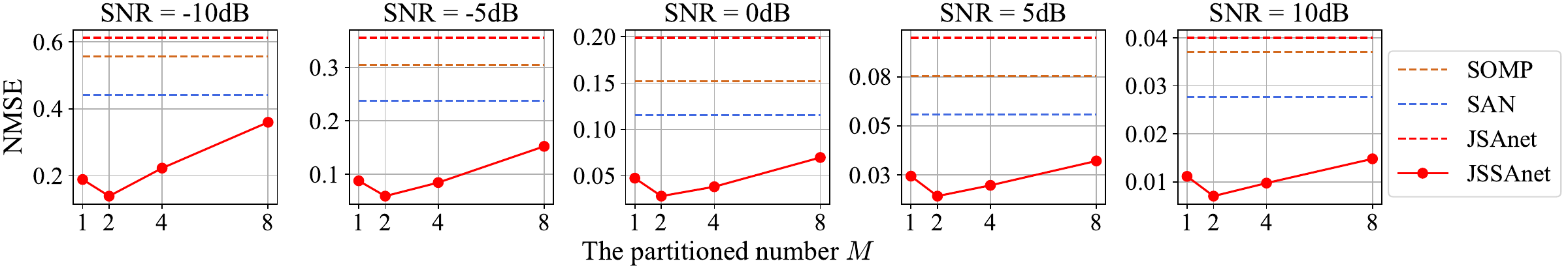}
	\caption{
		The NMSE performance of JSSAnet against $M$ at different SNRs.}
	\label{fig:subs_nmse}
\end{figure*}

\begin{figure*}[t] 
	\centering
	\includegraphics[width=0.95\textwidth]{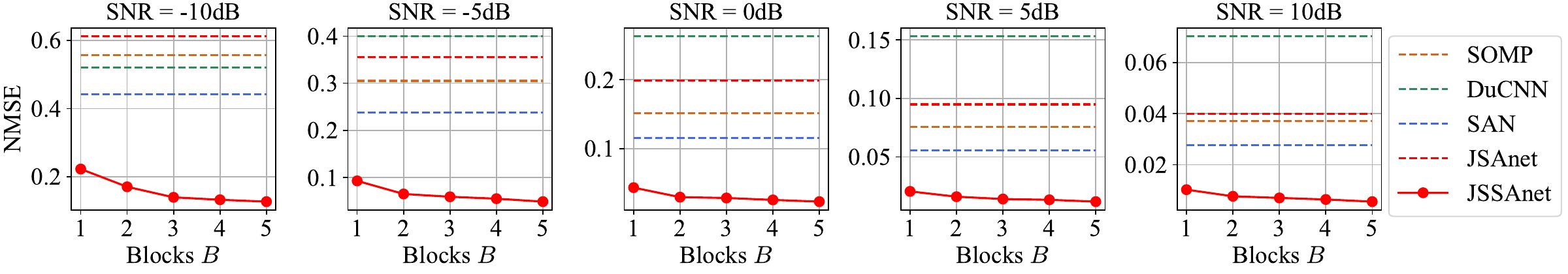}
	\caption{
		The NMSE performance of JSSAnet against JSSA blocks at different SNRs.}
	\label{fig:block_nmse}
	\vspace{-10pt}
\end{figure*}

\subsection{Performance across the Numbers of Partitioned Subchannels and Attention Blocks}

\begin{figure}[t]
	\centering
	\begin{subfigure}[t]{0.49\linewidth}
		\centering	
		\includegraphics[width=\textwidth]{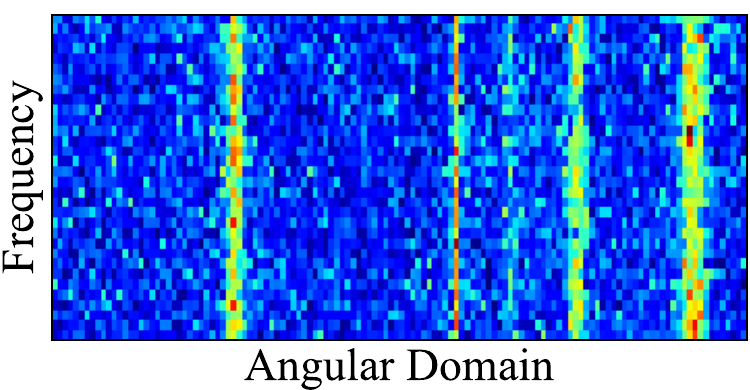}
		\caption{ Power spectrum of $\mathcal{D}(\mathbf{H}_1)$}
		\label{fig:subchannel1}
	\end{subfigure}
	\begin{subfigure}[t]{0.49\linewidth}
		\centering	
		\includegraphics[width=\textwidth]{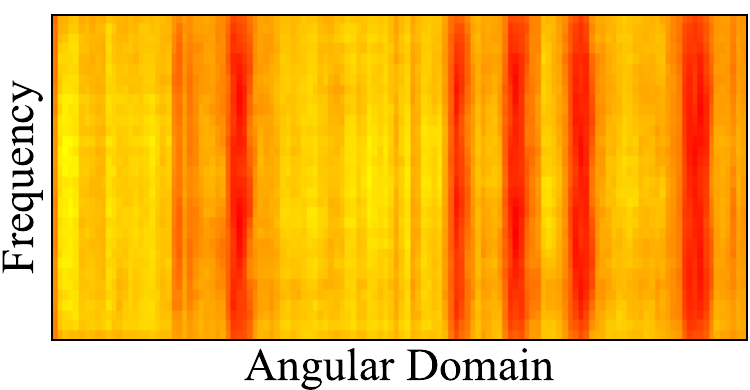}
		\caption{Attention map $\mathbf{M}_1$}
		\label{fig:subchannel1_hotmap}
	\end{subfigure}
	\begin{subfigure}[t]{0.49\linewidth}
		\centering	
		\includegraphics[width=\textwidth]{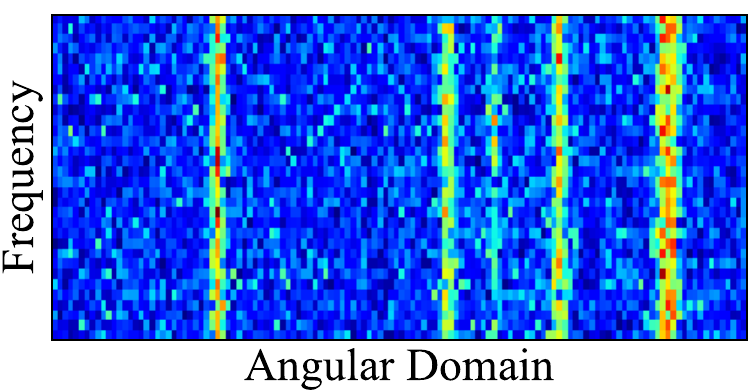}
		\caption{Power spectrum of $\mathcal{D}(\mathbf{H}_2)$}
		\label{fig:subchannel2}
	\end{subfigure}
	\begin{subfigure}[t]{0.49\linewidth}
		\centering	
		\includegraphics[width=\textwidth]{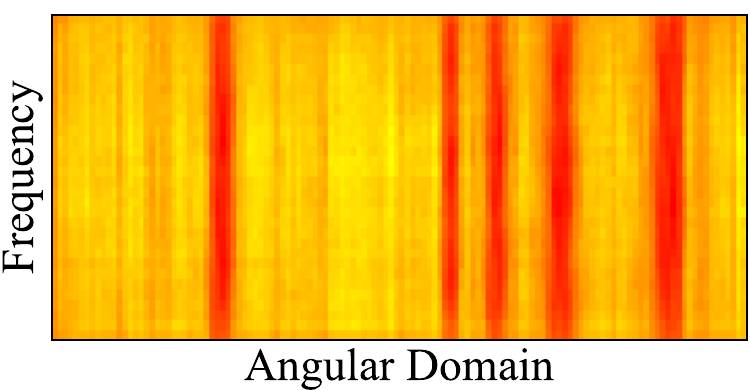}
		\caption{Attention map $\mathbf{M}_2$}
		\label{fig:subchannel2_hotmap}
	\end{subfigure}
	\caption{
		The power spectrum of subchannels in the angular domain and corresponding attention maps of the third JSSAB with SNR $=0$ dB and $r = 5$ m.
	}
	\label{fig:hotmap}	
\end{figure}
In order to experimentally validate the bounds of the partitioned number $M$ as stated in \textit{Theorem \ref{pos:M_min}} and \textit{Theorem \ref{pos:M_max}}, we compare a NMSE performance of our JSSAnet under $M \in \{1,2,4,8\}$ under different SNRs.
Specifically, the case of $M=1$ is that the angular representations of the near-field channels are fed into the JSSAnet.
As depicted in Fig. \ref{fig:subs_nmse}, it is evident that JSSAnet achieves the best NMSE performance when the antenna array is uniformly partitioned into $M=2$ subarrays.
The near-field channel suffers from angular energy spread so that JSSAnet with $M=1$ faces greater challenges in extracting inherent spatial properties of near-field channels than that with $M=2$.
As analyzed in Section \ref{subsection:number} and illustrated in Fig. \ref{fig:dft_sub4}, when $M>2$, the reduced spatial angular resolution leads angular information loss and renders DFT incapable of capturing inter-subchannel spatial-angular diversity.
Therefore,  it is highly challenging for JSSAnet to capture near-field  characteristics from multiple similar subchannel angular representations.
JSSAnet consistently outperforms other variants across different partitioned numbers, further validating the effectiveness of embedding channel spatial information into the attention mechanism for enhancing near-field channel estimation.

As presented in Fig. \ref{fig:block_nmse}, the performance of our JSSAnet with various JSSA blocks $B$ is evaluated.
We observe that NMSE performance improves until reaches a threshold as the number of blocks increases, particularly in the low SNR regime.
It is worthwhile noting that JSSAnet with $B=1$ maintains significantly better performance than other variants, further confirming the superiority of our JSSA mechanism in the near-field channel estimation.
While the increasing number of blocks enhances model capability, it also results in higher computational overhead.
In trade-off, $B=3$ blocks are adopted in our JSSAnet.

\begin{figure}[t]
	\centering
	\includegraphics[width=0.35\textwidth]{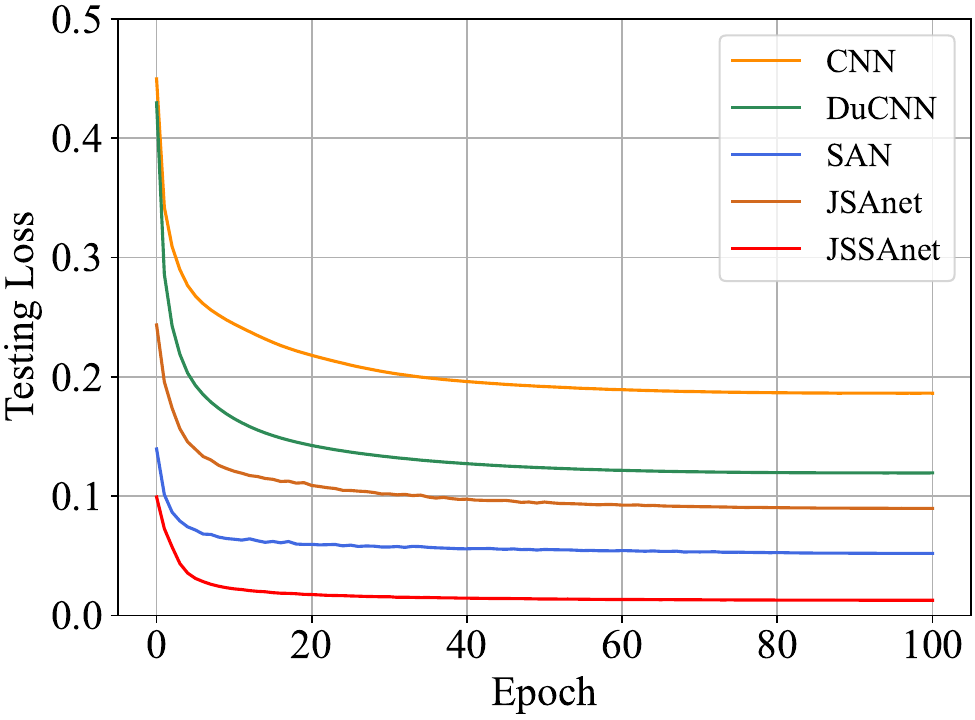}
	\caption{
		The convergence performance comparison of different network structures under SNR $=0$ dB.}
	\label{fig:loss}
\end{figure}

\subsection{Visualizations of Attention Maps and Loss Convergence}

For comprehensive analyze the processing of our JSSA mechanism, we provide visualizations of attention maps on angular representations of subchannels, as shown in Fig. \ref{fig:hotmap}.
Here, we choose the attention maps $\mathbf{M}_1$ and $\mathbf{M}_2$ from the third JSSAB.
Fig. \ref{fig:subchannel1} and \ref{fig:subchannel2} represent the power spectrum of $\mathbf{H}_1$ and $\mathbf{H}_2$ in the angular domain, respectively.
Fig. \ref{fig:subchannel1_hotmap} and \ref{fig:subchannel2_hotmap} depict the visualizations of attention map $\mathbf{M}_1 $ on $\mathcal{D}(\mathbf{H}_1)$ and attention map $\mathbf{M}_2 $ on $\mathcal{D}(\mathbf{H}_2)$, respectively.
$\mathbf{M}_1 $ and $\mathbf{M}_2 $ focus respectively distinct angular regions, as the angular beam patterns between subchannels are different.
It is evident that the JSSAB automatically allocates higher attention weights to spatial directions with stronger energy intensity, guided by the spatial angular beam patterns of the subchannels.

The testing loss over training epochs is evaluated, and the corresponding loss curves of the neural networks are shown in Fig. \ref{fig:loss}.
The testing loss gradually decrease as training progresses, indicating the parameters of the neural networks are being optimized during training period.
There is a fact that the attention-based schemes have lower losses at the beginning of training phase than CNN-based schemes.
Notably, the proposed JSSAnet has the lowest initial loss value and maintains consistently lower loss with the faster convergence throughout training phase.
Hence, it demonstrates the developed attention network achieves more efficient and accurate near-field channel estimation with linear computational complexity.

\section{Conclusion}\label{section:conclusion}

This paper proposed JSSAnet, that embeds the joint sparse angular information into the attention procedure, to improve the channel estimation in near-field communication scenario.
To address the overhead of expanded near-field dictionaries, we introduced the piecewise Fourier vector to implicitly encode the nonlinearity in the near-field ARV utilizing the diverse spatial angles of subarrays so that the channel is partitioned into subchannels, each of which is mapped to the angular domain via DFT.
To theoretically guide the design of the joint attention mechanism, we derived a theoretical lower bound of the piecewise number to satisfy the approximate criterion and an upper bound was established theoretically to preserve the angular diversity across subchannels after DFT projection.
In JSSAnet, each JSSA block was developed based on a decoupling-fusion strategy.
The JSSA layer firstly provides the personalized and independent spatial attentions for each subchannel in parallel, while FFN fuses the extracted subchannel features to model the nonlinear representations of near-field channel.
Notably, our JSSA mechanism adopts the element-wise product combining DLKC for linear complexity without compromising contextual learning capability.
Numerical simulations were provided to validate the proposed JSSAnet significantly outperforms other estimation schemes, especially DuCNN and SAN.
Ablation studies show JSSA mechanism improves significantly the accuracy of near-field channel estimation.

\appendices
\section{The Derivation of Theorem \ref{pos:M_min}} \label{app:1}
The similarity between the piecewise Fourier vector $\mathbf{b}(\bm{\theta})$ and near-field ARV $\mathbf{a}(\theta, r)$ is defined as $S = \mathbf{b}(\bm{\theta})^H\mathbf{a}(\theta_0,r_0)$.
Then, the similarity of the $i$-th subarray is expressed as
\begin{equation}
	\begin{aligned} 
		 S_i
		 =&  e^{-j\pi \tilde{p}_i} \mathbf{b}(\tilde{\theta}_i)^H [\mathbf{a}(\theta_0, r_0)]_{\mathcal{S}_i}
		 \\
		 =& \frac{1}{N_{\text{BS}}} \sum_{n=-\frac{N}{2}+1}^{\frac{N}{2}} e^{-j\pi \left (\tilde{p}_i+n\tilde{\theta}_i\right )} \cdot  e^{j\pi \left (\tilde{p}_i-\frac{1-\tilde{\theta}_i^2}{2\tilde{r}_i} d n^2 +n \tilde{\theta}_i \right )}
		 \\
		 \approx & \frac{2}{N_{\text{BS}}}  \sqrt{\frac{2\tilde{r}_i}{(1-\tilde{\theta}_i^2) d}} 
		\left [ C\left (\frac{N-1}{2} \sqrt{\frac{1-\tilde{\theta}_i^2}{2\tilde{r}_i} d }\right ) - \right. \\
		&  \left. j S\left (\frac{N-1}{2} \sqrt{\frac{1-\tilde{\theta}_i^2}{2\tilde{r}_i} d }\right ) \right ],
	\end{aligned}
\end{equation}
where the second exponential term of the second equality comes from the fact that the phase difference of the elements of the $i$-th subarray equals the sum of the phase difference between reference elements of the subarray and the ELAA and the phase difference  between elements and the reference element of th subarray;
and utilizing the Riemann sum approximate and the Fresnel functions yields the last approximate equality \cite{zhu_twc}.

We assume that $S_i\approx S_j$ for $i\neq j $, where $i,j\in \{1, 2,\cdots,M\}$.
Thus, we have $S = M S_i$.
The similarity metric $|S|$ can be asymptotically driven to approach $1$ as the partitioned number $M$ increases, enabling the piecewise Fourier vector to approximate the near-field ARV.

Invoking $C^2(X)+S^2(X)\leq 1$ \cite{Gradshteyn2014_Table} and making $|S|$ be larger than the 3 dB of the unit power, we set the coefficient meets
\begin{equation}
	\frac{2M}{N_{\text{BS}}}  \sqrt{\frac{2\tilde{r}_i}{(1-\tilde{\theta}_i^2) d}} \geq \frac{1}{\sqrt{2}}  \Rightarrow M \geq \frac{N_{\text{BS}}}{4}\sqrt{\frac{(1-\tilde{\theta}_i^2) d}{\tilde{r}_i}}
\end{equation}
to meet similarity metric.
Given $\min\left\lbrace \tilde{\theta}_i^2\right\rbrace=0$ and $\min\tilde{r}_i=r_\text{min} $, this theorem is proved.

\section{The Derivation of Theorem \ref{pos:M_max}} \label{app:2}
Referring to the power concentration property of $\Xi(x)$, the most of the power of $\mathbf{c}_{k,\ell}^{(i)} $ is focused on within bandwidth of $\frac{2}{N}$.
Namely, the spatial angular resolution of an $N\times N$ DFT matrix is equal to $\frac{2}{N}$.
Accordingly, the spatial AoA difference of inter-subarray must be larger than $\frac{2}{N}$ and the minimal spatial AoA difference of inter-subarray is geometrically constrained to occur between adjacent subarrays.
Thus, it is mathematically written by
\begin{equation}
	\min_{i \in\{1,\cdots, N-1\}} \left |\theta_\ell^{(i)}-\theta_\ell^{(i+1)}\right | \geq \frac{2}{N}.
\end{equation}
Invoking \eqref{eq:theta_taylor}, the spatial AoA difference between both adjacent subarrays is computed by $\frac{1-\bar{\theta}_i^2}{\bar{r}_i}d N$, where $\bar{\theta}_i$ and $\bar{r}_i$ are the spatial AoA and distance between the scatter and the center of both adjacent subarrays, respectively.
Given $M=\lceil \frac{N_\text{BS}}{N} \rceil$, we have 
\begin{equation}
	M \leq \min \sqrt{\frac{(1-\bar{\theta}_i^2)d}{2\bar{r}_i}} N_\text{BS}  .
\end{equation}
Given $\max \left\lbrace \bar{\theta}_i^2\right\rbrace  = \theta_\text{sec}^2 $ and substituting $r=r_\text{min} $, this theorem is proved.

\bibliography{ref}
\bibliographystyle{IEEEtran}
\end{document}